\documentclass[twocolumn, RNAAS]{aastex701}

\usepackage{booktabs}
\usepackage{amsmath}
\usepackage{subcaption}
\usepackage{hyperref}
\usepackage{graphicx}
\usepackage{txfonts}
\usepackage{lipsum}
\usepackage{cancel}
\usepackage{booktabs}
\usepackage{tabularx}
\usepackage[latin1]{inputenc}
\usepackage[T1]{fontenc}

\begin{document}

\title{No Significant Excess in Typical Variability in CLAS+ Changing-Look AGN Candidates in Rubin DP2}

\author[orcid=0000-0001-6480-1155]{L. Nakazono}
\affiliation{Departamento de F\'isica Matem\'atica, Instituto de F\'isica, Universidade de S\~ao Paulo,
R. do Mat\~ao 1371, 05508-090, S\~ao Paulo, SP, Brazil}
\affiliation{Observat\'orio Nacional/MCTI, Rio de Janeiro, RJ, Brazil}
\email[show]{liliannenakazono@on.br}

\begin{abstract}
Rubin Data Preview 2 (DP2) samples the CLAS+ changing-look AGN (CLAGN) candidates $\sim1,000-9,000$ days after their comparison epochs, so it can test whether these objects retain enhanced short-term optical variability after their spectral transitions. Comparing peak-to-peak amplitude in magnitudes against a photometrically quiescent quasar control, one band at a time, we do not detect a significant median-amplitude excess for any state (turn-on, turn-off, intermediate) in either of two bands, nor a different in the median $\,|\,\Delta \text{mag}\,|\,$ as function of time lag. The $r$-band CLAS+ amplitude distributions are wider than the control's, with the intermediate state formally significant after multiple-comparison correction. This could hint for a high-amplitude subpopulation. 
\end{abstract}



\section{Introduction}
CLAS+ selects changing-look AGN (CLAGN) candidates from a shift in S-PLUS (\citealt{2019MNRAS.489..241M}) narrow-band colors between up to two spectroscopic epochs, classified turn-on, turn-off, or intermediate\footnote{The intermediate state are a transitional phase in which the AGN shows spectral properties between its bright (turn-on) and faint (turn-off) states.}states \citep{nakazono2026claslargecatalogchanginglook}. Rubin/LSST \citep{2019ApJ...873..111I} offers the first chance to follow temporal variations of such a pre-selected sample, but DP2 is a brief early-commissioning dataset \citep{RTN-115}. Each field was visited in only a handful of campaigns spanning $\sim$1-3 months. The inclusion of an object in CLAS+ does not imply it should be undergoing a transition during DP2, which begins 1,123-9,443 days (median 1,385) after the most recent CLAS+ epoch. In this note, I instead investigated whether CLAS+ candidates still show excess short-term variability in this post-transition state.

\section{Data and Methods}
I crossmatch 3,263 CLAS+ objects against Rubin DP2 \texttt{DiaSource} photometry by position within 1\arcsec, obtaining 740 light curves (365 intermediate, 260 turn-on, 115 turn-off) after removing flagged epochs (1.7\%). As a control I identically crossmatch quasars with $\chi^2_r\sim1$ (no significant narrow-line change) from the same parent surveys, obtaining 225 control light curves after removing overlaps. I use total flux (\texttt{scienceFlux}).

I restrict to the $r$ and $i$ bands and $\geq 5$ epochs, leaving 115 (30) in r and 134 (36) in i CLAS+ (control) objects; u, z and y leave too few objects, and g shows an $\sim0.12$ mag amplitude floor shared by both samples, a processing artifact rather than intrinsic variability.  The $r$-band subsample skews to lower redshift than the parent CLAS+ sample (median z 1.11 vs. 1.31; KS $p=0.005$); the $i$-band subsample does not. Three test families are defined in advance: 6 median tests (3 states $\times$ 2 bands), 6 dispersion tests, and 33 structure-function lag bins; with $p$-values corrected within each.

\section{Results}
\begin{figure*}
\includegraphics[width=1\textwidth, trim= 0cm 0cm 0cm 1cm, clip]{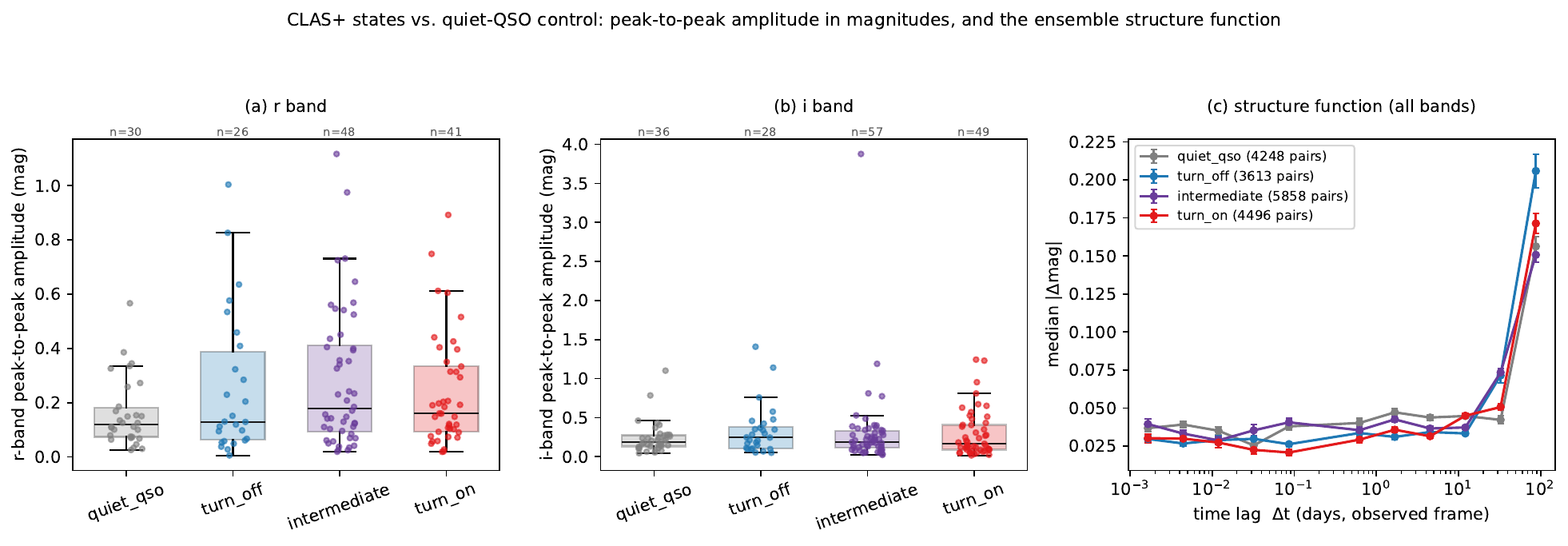}
\caption{CLAS+ changing-look candidates vs. the quiescent quasar control in Rubin DP2. (a, b) Peak-to-peak scienceFlux amplitude in magnitudes for objects with $\geq$ 5 epochs in r (a) or i (b); boxes span the quartiles with the median marked, points are individual objects, n per group above each box. No state's median differs significantly from the control in either band. The r-band CLAS+ distributions are broader; only the intermediate state's spread survives multiple-comparison correction, concentrated in a high-amplitude tail. (c) Median $\,|\,\Delta mag\,|\,$ between within-light-curve epoch pairs vs. observed-frame time lag, one curve per state. No lag bin differs significantly from the control; the long-lag rise, present in the control too, reflects DP2 cross-campaign cadence rather than intrinsic growth.}
\label{fig:1}
\end{figure*}

For each object I compute the peak-to-peak amplitude of \texttt{scienceFlux} per band in magnitudes, testing within a single band to avoid a bias from CLAS+ having more sampled bands than the control. No state reaches significance in either band (Fig. \ref{fig:1} : $r$, $p$ = 0.13/0.50/0.065; $i$, $p$ = 0.78/0.26/0.98; turn-on/turn-off/intermediate vs. control, Mann-Whitney). Effect sizes are small (rank-biserial r $\leq$ 0.25), and the smallest $p$ (intermediate, r) rises to 0.39 after Holm-Bonferroni correction. This agrees with the median $\,|\,\Delta mag\,|\,$ vs. time lag (Fig. \ref{fig:1}c): binning within-light-curve epoch pairs into 12 log-spaced lag bins and taking each object's median $\,|\,\Delta mag\,|\,$ per bin, no populated (state $\times$ bin) test reaches significante (minimum raw $p =$ 0.09; Benjamini-Hochberg-adjusted $p \geq 0.88$), in either observed or rest-frame lag.

The medians are consistent,  but $r$-band CLAS+ amplitude distributions are wider than the control's (standard-deviation ratio $1.6-2.1\times$ across states). After Holm-Bonferroni correction, the $r$ intermediate state remains significant (Brown-Forsythe $p=0.04$, Fligner-Killeen $p=0.01$); turn-on/turn-off show the same effect size but do not survive correction, and no $i$-band state does. This is not a baseline-mismatch artifact as restricting to \texttt{mjd\_span} $\geq20$ days balances the baseline distributions and leaves the $r$ intermediate result intact (Brown-Forsythe $p=0.004$). The difference concentrates in the upper tail: 40\% of $r$ intermediate light curves exceed $\,|\,\Delta mag\,|\,=0.3$ (vs. 17\% of control; Fisher $p=0.04$). With one state surviving correction, in one band, on $n\approx 13-48$, and no luminosity match, I treat this as a hint at a high amplitude tail to follow up. 

Amplitude is compared in magnitudes rather than raw flux because CLAS+ is 1.8-2.6$\times$ brighter than the control. A flux-based amplitude reproduces this brightness difference as a spurious excess that vanishes once magnitudes are used. Per-epoch noise, larger for the fainter control (median SNR $\sim$165 vs. $\sim$280), works against rather than for a CLAS+ excess, and the $r$ turn-on-vs-control test ($n=41$ vs. 30) has only $\sim$80\% power to detect a $\sim$0.12 mag mean difference, so smaller true excesses could go undetected.

The samples are not matched in luminosity. Using DESI DR1 single-epoch Mg II measurements \citep{2025JCAP...01..124A} for the subset that has them, CLAS+ is $\sim$0.25 dex more luminous and $\sim$0.3 dex more massive in black hole than the control, with an indistinguishable Eddington ratio. Since variability generally anticorrelates with luminosity \citep{MacLeod_2010}, this would tend to work against a CLAS+ excess, but the net effect is undemonstrated. A luminosity-, redshift-, and cadence-matched control would settle this.

\section{Conclusion}
This is the first Rubin photometric look at the CLAS+ sample. In the $\sim$1-3 month DP2 window CLAS+ candidates are not detectably more variable than a quiescent control in median amplitude or structure function. The one difference that survives multiple-comparison correction is a broader $r$-band amplitude distribution for the intermediate state, from a high-amplitude minority. Overall, DP2 is too short to show how changing-look variability evolves after a transition. The multi-year LSST baseline would test that directly. 
\begin{acknowledgments}
This material is based upon work supported in part by the National Science Foundation through Cooperative Agreements AST-1258333 and AST-2241526 and Cooperative Support Agreements AST-1202910 and 2211468 managed by the Association of Universities for Research in Astronomy (AURA), and the Department of Energy under Contract No. DE-AC02-76SF00515 with the SLAC National Accelerator Laboratory managed by Stanford University. Additional Rubin Observatory funding comes from private donations, grants to universities, and in-kind support from LSST-DA Institutional Members.
This publication is based in part on proprietary Rubin Observatory Legacy Survey of Space and Time (LSST) data, and was prepared in accordance with the Rubin Observatory data rights and access policies. All authors of this publication meet the requirements for co-authorship of proprietary LSST data.

This research made use of NumPy \citep{harris2020array}, pandas
\citep{reback2020pandas}, SciPy \citep{2020SciPy-NMeth}, Matplotlib \citep{Hunter:2007},
and Astropy \citep{2022ApJ...935..167A}. Anthropic's Claude was used to develop the analysis pipeline, and draft and revise the manuscript. The author verified all outputs and takes full responsibility for the work.

\end{acknowledgments}

\bibliography{ApJ}{}
\bibliographystyle{aasjournalv7}



\end{document}